\documentclass[showpacs,onecolumn,amsmath,nofootinbib,amssymb,natbib]{revtex4}
\usepackage{graphicx,epsfig,dcolumn,bm}
\usepackage{epsfig}
\usepackage{float}

\newcommand{\ba}{\begin{eqnarray}}
\newcommand{\ea}{\end{eqnarray}}
\newcommand{\bege}{\begin{equation}}
\newcommand{\enge}{\end{equation}}
\newcommand{\benu}{\begin{enumerate}}
\newcommand{\enu}{\end{enumerate}}
\newcommand{\pa}{\partial}
\newcommand{\noi}{\noindent}
\newcommand{\bbbbox}{\mathop{\Box\kern -5pt\raisebox{.8pt}{$|$}}}

\newcommand{\mr}{\mathring}

\newcommand{\OO}{\mathbb{O}}

\def\beq{\begin{eqnarray}}
\def\eeq{\end{eqnarray}}

\def\ua{\{-,+\}}
\def\da{\{+,-\}}

\def\ua{\{-,+\}}
\def\da{\{+,-\}}

\def\p{\mbox{\boldmath$\displaystyle\mathbf{p}$}}

\def\0{\mbox{\boldmath$\displaystyle\mathbf{0}$}}

\def\h00h{\mbox{\boldmath$\displaystyle\mathbf{(1/2,0)\oplus(0,1/2)}$}}

\def\ba{p \,e^{i\phi}  \sin(\theta)}

\def\beq{\begin{eqnarray}}
\def\eeq{\end{eqnarray}}

\def\ua{\{-,+\}}
\def\da{\{+,-\}}

\def\ua{\{-,+\}}
\def\da{\{+,-\}}

\def\p{\mbox{\boldmath$\displaystyle\mathbf{p}$}}

\def\0{\mbox{\boldmath$\displaystyle\mathbf{0}$}}

\def\h00h{\mbox{\boldmath$\displaystyle\mathbf{(1/2,0)\oplus(0,1/2)}$}}

\def\ba{p \,e^{i\phi}  \sin(\theta)}

\begin{document}

\title{Dynamical dispersion relation for ELKO dark spinor fields}
\author{A. E. Bernardini}
\email{alexeb@ufscar.br, alexeb@ifi.unicamp.br}
\affiliation{Departamento de F\'{\i}sica, Universidade Federal de S\~ao Carlos, PO Box 676, 13565-905, S\~ao Carlos, SP, Brasil}
\author{Rold\~ao da Rocha}
\email{roldao.rocha@ufabc.edu.br}
\affiliation{Centro de Matem\'atica, Computa\c c\~ao e Cogni\c c\~ao,
Universidade Federal do ABC 09210-170, Santo Andr\'e, SP, Brazil}


\begin{abstract}
An intrinsic mass generation mechanism for exotic ELKO dark matter fields is scrutinized, in the context of the very special relativity (VSR).
Our results are reported on unraveling inequivalent spin structures  that educe an additional term on the associated Dirac operator. Contrary to the spinor fields of mass dimension 3/2, this term is precluded to be absorbed as a shift of some gauge vector potential, regarding the equations for the dark spinor fields.
It leads to some dynamical constraints that can be intrinsically converted into a dark spinor mass generation mechanism, with the encoded symmetries maintained by the VSR.
The dynamical mass is embedded in the VSR framework through a natural coupling to the {\em kink} solution of a $\lambda \phi^{4}$ theory for a scalar field $\phi$.
Our results evince the possibility of novel effective scenarios, derived from exotic couplings among dark spinor fields and scalar field topological solutions.

\end{abstract}

\maketitle

\section{Introduction}

The investigation on the nature of dark matter components, as well as the comprehension regarding all their intrinsic relations with the elements of the cosmic inventory, belong to one of  challenging current problems in theoretical physics \cite{Teo02,Teo03,Teo04,Teo01}.
A novel form of matter called ELKO, the acronym of {\em Eigenspinoren des Ladungskonjugationsoperators}, which designates the eigenspinors of the charge conjugation operator, seems to fulfill the requirements for a dark matter component, in the scope of the interplay among general relativity, astrophysics and particle physics \cite{Roc11,boe1,boe2,boe3,boe4,fabbri1,fabbri2,fabbri3,where,where1,where2,where3}. These references for instance evince that ELKO spinor fields
 main interaction via the gravitational field makes them naturally dark, which inforces dark spinor fields investigation in a cosmological setting, where interesting solutions and also models where the spinor is coupled conformally to gravity are provided. 
Once embedded in the quantum field theory, besides leading to some non-local properties \cite{allu,allu1,allu2,allu3}, the standard formulation of ELKO predicts modified dispersion relations. Furthermore, it allows for accomplishing dual-helicity mass dimension one eigenspinors of the spin-$1/2$ charge conjugation operator.
The possibility of exotic interactions with the Higgs scalar field, and suppressed interactions with gauge fields, accredits such matter fields as potential candidates to describe dark matter \cite{allu,allu1,allu2,allu3}.
At standard model (SM) energy scales, ELKO should behave as a representation of the Lorentz group through the setup of a preferred direction related to its wave equation \cite{allu,allu1,allu2,allu3,alex,alex1,horv1}.
It is recovered by the conjecture of the very special relativity (VSR) \cite{glashow,glashow1}.
The Lorentz symmetries underlying the SM matter and gauge fields, as well as the algebraic structure underlying VSR \cite{horv1} supported by the event space underlying dark matter and dark gauge fields, have been continuously evaluated,  in order to describe the embedding of dark spinor fields into the SM \cite{osmano,osmano1}.

The VSR operates at the Planck scale to reproduce the SM as an associated effective theory.
It is supposed to be operative not solely at ultra-high energies, but also beyond SM energy scales, where  dark matter interactions may eventually take place.
In the context of elementary interactions between fermonic and gauge fields, as well as to preserve the intrinsic darkness with respect to the SM gauge fields \cite{horv1}, it is possible to construct a VSR invariant fermionic field with unitary mass dimension and spin-$1/2$: the ELKO.

To shed some primordial light on ELKO dark matter field properties, one may reckon that in spacetimes with non-trivial topology there ought to be an additional degree of freedom for fermionic particles \cite{Roc11, Asselmeyer:1995jp,isham1,isham2,isham3,isham4}.
Such novel element emerges when, for instance, SM spinor fields are parallel transported: a complementary one-form field $\xi^{-1}(x) d\xi(x)$
is accrued on the Dirac operator, educed by the non-trivial topology \cite{Roc11, Asselmeyer:1995jp,isham1,isham2,isham3,isham4}. Here  $d$ denotes the exterior derivative operator and $\xi$ is a scalar field.
 When SM Dirac spinor fields are taken into account, the vector gauge field $V$ term is affected by the transformation $V \mapsto V + \frac{1}{2\pi i} \xi^{-1} d\xi$, which indeed corresponds to the addition of an  gauge potential extra term.
Such an exotic term may be therefore absorbed by an external abelian gauge potential, representing an element of the cohomology group $H^1(M, \mathbb{Z}_2)$ \cite{isham1,isham2,isham3,isham4}, in the context of the (exotic) Dirac equation. Notwithstanding, concerning exotic ELKO dark spinor fields the possibility of such an intrinsic coupling to SM gauge fields is null. Since gauge fields interactions with ELKO are suppressed \cite{allu,allu1,allu2,allu3}, ELKO fields are able to   probe purely the space-time topology \cite{Roc11}.
It consistently reinforces the above mentioned darkness of such fields.
This paper is organized as follows: in the next Section,
the exotic structure underlying ELKO dark spinor fields \cite{Roc11} is briefly revisited. In Section \ref{3se} we identify some similarities between the exotic  Dirac operator and the covariant derivative embedded into the framework of the VSR, so that the Klein-Gordon equation for a massive particle can be reproduced.
By identifying the VSR preferential direction with a dynamical dependence on the {\em kink} solution of a $\lambda \phi^{4}$ theory for a scalar field $\phi$, we show that an effective mass for the ELKO spinor can be naturally obtained.
It evinces the possibility of novel scenarios for the mechanism of dynamical mass generation, as well as for exotic couplings with scalar field topological solutions.
In Section \ref{4} we conclude and provide novel perspectives on the potentially relevant and prominent results addressed in this paper.

\section{Exotic ELKO dark spinor fields}

ELKO spinor fields $\lambda({\bf{p}})$ are defined as eigenspinors of the charge conjugation operator $C=\scriptsize{\begin{pmatrix}
\OO & i\Theta \\
-i\Theta & \OO \nonumber
\end{pmatrix}}K$, in the precise sense that $
C\lambda({\bf{p}})=\pm \lambda({\bf
p})$,
where,  given the rotation generators
${\mathcal{J}}$, the Wigner's spin-1/2 time reversal operator
$\Theta$ satisfies $\Theta
\mathcal{J}\Theta^{-1}=-\mathcal{J}^{\ast}$. The operator $K$ is responsible to $\mathbb{C}$-conjugate  spinor fields appearing on the right.
The plus [minus] sign stands for self-conjugate [anti self-conjugate] spinor fields $\lambda^{S}({\bf p})$ [$\lambda^{A}({\bf p})$].
The complete form of ELKO can be explicitly obtained \cite{allu,allu1,allu2,allu3} through the solution of the equation of helicity $(\sigma\cdot\widehat{\bf{p}})\lambda^{A/S}_\pm({\bf p})=\pm \lambda^{A/S}_\pm({\bf p})$, where $\widehat{\bf{p}}={\bf p}/\|{\bf p}\|$. The four boosted spinor fields are\footnote{The
boosts here presented are Lorentz boosts, although SIM(2) VSR boosted ELKO can be derived as in \cite{horv1}. For our aim in what follows Lorentz boosts suffice, and therefore shall be adopted. }
\begin{equation}
\lambda^{S/A}_{\{\mp,\pm \}}({\bf
p})=\sqrt{\frac{E+m}{2m}}\Bigg(1\mp
\frac{p}{E+m}\Bigg)\lambda^{S/A}_{\{\mp,\pm \}}({\bf{0}}),\qquad{\rm  where}\qquad\lambda^{S/A}_{\{\mp,\pm \}}(\bf{0})=%
\begin{pmatrix}
\pm i \Theta[\phi^{\pm}(\bf{0})]^{*} \\
\phi^{\pm}(\bf{0})\label{five}\nonumber
\end{pmatrix}
\end{equation}
The (Weyl) spinors fields
$\Theta[\phi^{\pm}(\bf{0})]^{*}$ and $\phi^{\pm}(\bf{0})$ have opposite helicities.

Prior results moreover evince that ELKO can be expressed through a linear combination of Dirac particle and antiparticle fields \cite{allu,allu1,allu2,allu3,where,where1,where2,where3,Roc11}, and the prescription $p_\mu\mapsto i \nabla_\mu$ holds for ELKO:  $\lambda^\mathrm{S/A}(x)=\lambda^\mathrm{S/A}(\p)
\;\exp\left(\epsilon^\mathrm{S/A}\;\,i p_\mu x^\mu\right)$, where $\epsilon^\mathrm{S} = -1$ and $\epsilon^\mathrm{A} = +1$  \cite{allu}.

Besides the standard ELKO spinor fields $\lambda(x)$,
one can get a second type of ELKO, denoted hereupon by $\mathring\lambda(x)$, associated to an inequivalent spin structure, that reflects  a modification of the covariant derivative \cite{isham1,isham2,isham3,isham4}:
\bege
\mr{\nabla}_X{\mathring\lambda(x)} = \nabla_X{\mathring\lambda(x)} -\frac{1}{2}\left[X\cdot\left(\xi^{-1}(x)d\xi(x)\right)\right]\mathring\lambda(x)\label{covvv},
\enge where $X$ denotes a vector field.
The so called exotic term in Eq.(\ref{covvv}) is assumed to be re-scaled as $\frac{1}{2\pi i}\left(\xi^{-1}(x)d\xi(x)\right)$,  an integer of a ${\check{\rm C}}$ech cohomology class \cite{Roc11,isham1,isham2,isham3,isham4}.
The exotic Dirac operator can be written thereupon as
\begin{eqnarray}
 i\gamma^\mu\mr\nabla_\mu = i\gamma^\mu\nabla_\mu + \frac{1}{2}\xi^{-1}(x)d\xi(x),\label{mrt}
\end{eqnarray}
\noi and the exotic Dirac equation then becomes
\[(i\gamma^\mu \nabla_\mu + (\xi^{-1}(x)\,d\xi(x))/2 \pm m\mathbb{I}) \psi(x) = 0,\]
\noi where $\psi$ denotes a Dirac spinor field.
As sustained in \cite{Asselmeyer:1995jp,isham1,isham2,isham3,isham4}, one can express $\xi(x) = \exp(2i\theta(x))\in$ U(1), so that the exotic term yields $
\xi^{-1}(x) d\xi(x) = 2i\gamma^\mu\partial_\mu\theta(x)$. One hence obtains the explicit form for the coupled system of  equations for the exotic ELKO  as
\beq
\left((i\gamma^\mu \nabla_\mu + i\gamma^\mu\pa_\mu\theta) \delta_\alpha^\beta \pm m\mathbb{I}
\varepsilon_\alpha^\beta\right)\mathring\lambda_\beta^\mathrm{S/A}(x)=0\,,\qquad \varepsilon^{\{-,+\}}_{\{+,-\}}:=-1\label{elko3}
\eeq\noi or, more explicitly --- taking into account Eq.(\ref{mrt}):
\beq
\hspace*{-1.2cm}\begin{pmatrix}
 i\gamma^\mu\mr\nabla_\mu& \mathbb{O} & \mathbb{O} & \mathbb{O} \\
 \mathbb{O} & i\gamma^\mu\mr\nabla_\mu& \mathbb{O} & \mathbb{O}  \\
  \mathbb{O} & \mathbb{O}& i\gamma^\mu\mr\nabla_\mu& \mathbb{O}   \\
  \mathbb{O} & \mathbb{O} & \mathbb{O} & i\gamma^\mu\mr\nabla_\mu \\
\end{pmatrix}
\begin{pmatrix}
\mathring\lambda^\mathrm{S}_{\ua} \\
\mathring\lambda^\mathrm{S}_{\da} \\
\mathring\lambda^\mathrm{A}_{\ua} \\
\mathring\lambda^\mathrm{A}_{\da}
\end{pmatrix}
-
i m  \mathbb{I}
\begin{pmatrix}
- \,\mathring\lambda^\mathrm{S}_{\da} \\
 \,\mathring\lambda^\mathrm{S}_{\ua} \\
 \,\mathring\lambda^\mathrm{A}_{\da} \\
- \,\mathring\lambda^\mathrm{A}_{\ua}
\end{pmatrix} = 0. \nonumber
\eeq
The exotic operator
$i\gamma^\mu {\nabla}_\mu + i\gamma^\mu\pa_\mu\theta \pm  m \mathbb{I}$
annihilates each of the four exotic Dirac spinor fields used to construct $\mathring\lambda_\beta^\mathrm{S/A}(x)$, as prescribed by the standard Dirac dynamics.
However, since the operator in (\ref{elko3}) couples the $\scriptsize{\{\scriptsize\pm,\scriptsize\mp\}}$ degrees of freedom \cite{allu,Roc11},
the modified exotic Dirac operator does not annihilate ELKO fields.
By observing the off-diagonal nature of the mass term in Eq.(\ref{elko3}) one should notice the differences from a phenomenological off-diagonal Majorana mass term introduced in the context of the Dirac equation \cite{allu,Roc11}.

Since the prerogatives for the ELKO dynamics are established \cite{allu,Roc11}, we drive our attention to the procedure for obtaining the corresponding effective dispersion relation derived from the exotic Dirac operator.
By analogy with the relativistic quantum mechanics terminology, we shall discuss whether the exotic Dirac operator can be considered as a square root of the Klein\textendash Gordon operator
--- in the sense that
$(i\gamma^\mu\nabla_\mu+i\gamma^\mu\pa_\mu\theta-m\mathbb{I})(i\gamma^\mu\nabla_\mu+i\gamma^\mu\pa_\mu\theta+m\mathbb{I}
)=(g^{\mu\nu}\nabla_\mu \nabla_\nu + m^2)\mathbb{I}$. This feature must remain true for the 5LKO and its exotic partner:
\beq\label{propa}\hspace*{-0.5cm}((i\gamma^\mu\nabla_\mu +i\gamma^\mu\pa_\mu\theta)\delta_\alpha^\beta \pm m\mathbb{I}
\varepsilon_\alpha^\beta)((i\gamma^\mu\nabla_\mu+i\gamma^\mu\pa_\mu\theta)
\delta_\alpha^\beta\mp  m\mathbb{I}
\varepsilon_\alpha^\beta)=(g^{\mu\nu}\nabla_\mu \nabla_\nu + m^2)\mathbb{I}\,
\delta_\alpha^\beta,\eeq\noi since the introduction of an exotic spin structure
does not modify the Klein\textendash Gordon propagator fulfillment by dark spinor fields.
The corresponding Klein-Gordon equation for the exotic ELKO field is hence provided by
\bege\label{exo1}
(\Box + m^2 +g^{\mu\nu}\nabla_\mu\nabla_\nu\theta + \pa^\mu\theta\nabla_\mu + \pa^\mu\theta\pa_\mu\theta)\mathring\lambda(x)^{S/A}_{\{\pm,\mp\}} = 0.\nonumber\enge
This equation can reproduce the same Klein-Gordon propagator for standard and exotic ELKO as well. For the exotic case, it demands the constraint
\bege
(\Box\theta + \pa^\mu\theta\nabla_\mu + \pa^\mu\theta\pa_\mu\theta)\mathring\lambda^{S/A}_{\{\pm,\mp\}}(x)=0\label{kg1}
\enge
which can be formulated without restricting the theory to any particular condition as those assumed in \cite{Roc11}.

\section{The exotic term in the VSR framework}\label{3se}

Instead of assuming the constraint given by Eq.~(\ref{kg1}) howsoever, one can suppress the effective mass term $m$ from the formalism, by assuming the exotic Dirac operator 
\bege
i\gamma^\mu\mr\nabla_\mu = i\gamma^\mu(\nabla_\mu + \partial_\mu \theta).
\label{theta}
\enge
Accordingly, it could be straightforwardly embedded into the framework of the VSR, when the exotic term $\partial_{\mu}\theta(x)$ is identified with a dynamical preferential direction vector $\phi(x) v_{\mu}$, where $\phi(x)$ is a scalar field with mass-dimension equal to one.
Thereafter obtaining the (preliminarily) massless Klein-Gordon equation from Eq.(\ref{theta}) as
\bege\label{KGKG}
(\Box + g^{\mu\nu} v_{\mu} v_{\nu} \phi^2  + g^{\mu\nu} v_{\mu} \partial_{\nu} \phi(x) + v^\mu \phi\nabla_\mu) \mathring\lambda(x)^{S/A}_{\{\pm,\mp\}} = 0,
\enge
one can investigate some constraints on the preferential vector $v_{\mu}$.

In order to shed some new light on the character of the exotic function $\theta(x)$ in terms of $\phi(x)$, let us identify the scalar field $\phi$ with the {\em kink} solution of the $\lambda \phi^{4}$ theory.
The {\em re-scaled} Lagrangian density for the scalar field of the $\lambda \phi^{4}$ theory has the form
\beq
\frac{1}{2} \partial_\mu\phi \partial^\mu\phi - s^2\frac{\lambda}{4}\left(\phi^{2} - \frac{m^2}{\lambda}\right)^{2},
\eeq
for which $m$ is the mass of the scalar field, $\lambda$ denotes the dimensionless coupling parameter, and $s^2 = g_{\mu\nu} s^{\mu}s^{\nu}$, with $s^{\mu} = v^{\mu} \sqrt{\frac{2}{\lambda}} = a^{\mu\nu}u_{\nu}$. The tensor  $a^{\mu\nu}$ is assumed to be anti-symmetric and $u_{\mu}$ is an unitary quadri-velocity ($u^2 = g_{\mu\nu} u^{\mu}u^{\nu} = 1)$.
The corresponding equation of motion yields
\beq
\Box\phi = \left(\frac{d^{2}}{dt^2} - \nabla^{2}\right)\phi = s^2 (m^2 \phi - \lambda \phi^3 ),
\eeq
and the  re-scaled \emph{kink} solution can be written as
\beq
\phi_s(x) = \frac{m}{\sqrt{\lambda}} \tanh\left(\frac{m}{\sqrt{2}} q \right), ~~\mbox{with} ~~ q = s^{\mu} x_{\mu}.
\eeq
Substituting the {\em kink} solution above into the {\em ansatz} relation $\partial_{\mu} \theta(x) = \phi(x) v_{\mu}$, one forthwith finds that
\beq
\theta(x) = \log\left[\cosh{\left(\frac{m}{\sqrt{2}} q \right)}\right],
\eeq
Therefore one can compute immediately
\beq
\hspace*{-0.2cm}\partial^{\mu}\theta\partial_{\mu}\theta + \Box\theta  &=& \frac{m^2}{2} s^2 \tanh[q^2] + \frac{m^2}{2} s^2 (1 - \tanh[q^2])\nonumber\\ &=& \frac{m^2}{2} s^2
\label{mass}
\eeq
Since $h^{\mu\nu}$ is an anti-symmetric tensor, one should notice that $u_{\nu}h^{\mu\nu}u_{\mu} = 0$.
One can thus parameterize the ELKO spinor field quadri-momentum $p_\mu$ by $p_{\mu} = \mu u_{\mu}$ in order to have
\beq
\partial_\mu \theta(x) \nabla^{\mu} \mathring\lambda^{S/A}_{\{\pm,\mp\}}(x) \propto  u_{\nu}h^{\mu\nu}u_{\mu} = 0.
\label{mass2}
\eeq
The results from Eqs.~(\ref{mass}-\ref{mass2}) substituted into Eq.~(\ref{KGKG}) lead to
\beq
\left(\Box + \frac{m^2}{2} s^2\right) \mathring\lambda^{S/A}_{\{\pm,\mp\}}(x)  = 0,
\label{mass3}
\eeq
which is typically the KG equation for a massive particle for which the dispersion relation would be given by $p^{\mu}p_{\mu} = (m^{2} s^{2})/2$.
The  result above eliminates the constraint given by Eq.~(\ref{kg1}) and reproduces the dynamics of the KG equation with a dynamical mass given by $\mu = m s/\sqrt{2}$.

\section{Concluding remarks}\label{4}

The VSR dynamical mass generation mechanism that we have suggested exactly reproduces the dispersion relations of SM fields, making the mechanism consistent with the VSR proposal.
It indeed emulates the conditions for generating dynamical masses that emerge in the context of dark matter coupling to dark energy intermediated by scalar fields \cite{Ber08A,Ber08B,Ber08C,Ber08D}.
Obviously, additional and natural questions may arise in a systematic formulation of quantum field theories carrying VSR symmetries \cite{horv1}.
The main outstanding exotic dark spinor field feature is that it can be indeed  considered in a variety of problems, wherein SM spinor fields cannot.
Observational aspects on such a possibility have been proposed at LHC, where  ELKO dark matter fields signatures can be elicited, at center of mass energy around 7 TeV. It robustly indicates the number of events that stimulates more specific analysis about the ELKO particle at high energy experiments \cite{marcao}. ELKO dark matter fields are manifestly non-local, although they present locality when the field propagation is along the preferred axis \cite{allu,horv1,local,local1}. Such axis can be chosen as the dynamical preferential direction $v_\mu$, provided by the exotic term. Innate difficulties to formulate a complete QFT for ELKO may be circumvented in our formalism, since such choice makes the ELKO a local field.

Any attempt to construct an appropriate QFT describing ELKO fermionic fields must accomplish their unexpected and interesting properties: non-locality, the existence of an inherent preferred axis (along which ELKO is local),  mass dimension one, as well as the Lorentz symmetry-breaking provided by the ELKO dynamics \cite{allu,where,where1,where2,where3,boe1,boe2,boe3,boe4,fabbri1,fabbri2,fabbri3,horv1,local,local1}.
The VSR has been evinced as a paramount promising arena to develop such still lacking  theory \cite{horv1}. Furthermore, such formalism can be investigated in a Hopf algebras scenario,  considering a state-space with symmetries of the Poincar\'e group. One uses the deformed Hopf algebras framework to construct an event-space with symmetries of the SIM(2) VSR group, in close analogy to the results already accomplished for the E(2) VSR group \cite{ture,ture1}.

The  ELKO dark spinor fields characterization is obtained through the topologically impelled introduction of an additional term  $i\gamma^{\mu}\partial_{\mu}\theta(x)$ in the Dirac operator, that cannot be absorbed by any external gauge field.
Even acting effectively in producing the exotic dispersion relation that we have obtained, modifications like $i\gamma^{\mu}\partial_{\mu}\theta(x) \mapsto i\gamma^{\mu}\partial_{\mu}(\theta(x) + \beta(x))$ with $\partial_{\mu}\theta(x)\partial^{\mu}\beta(x) = 0$, besides contributing to the dynamical mass term, can provide the preliminary conditions for the constraint on the metric spacetime structure, that emerges through the physical assumption that the exotic dark spinor fields satisfy the Klein-Gordon propagator for massive particles.
Therefore, our results heretofore assert that all the quantum field theoretical structure, that seems to be most suitable for describing dark matter, are maintained through the constraints previously applied to $\theta(x)$, when they are applied to $\beta(x)$.
Reverberating some previous conclusions, it means that the VSR reflects symmetries associated with rods and clocks for dark matter in the same way that theory of special relativity does for SM fields \cite{horv1,clo}.

\section*{Acknowledgment}

{
R. da Rocha is grateful to Conselho Nacional de Desenvolvimento Cient\'{\i}fico e Tecnol\'ogico (CNPq) grants {476580/2010-2} and
304862/2009-6 for financial support. A. E. B. would like to thank the financial support from CNPq (grant 300233/2010-8). The Authors are grateful to Prof. D. V. Ahluwalia,
for his fruitful suggestions.}


\begin{thebibliography}{99}
\bibitem{Teo02}
Y. B. Zeldovich, \emph{Gravitational instability: An Approximate theory for large density perturbations, Astron. Astrophys.} {\bf 5} (1970) 84.
\bibitem{Teo03}
L. Bergstrom, \emph{Nonbaryonic dark matter: Observational evidence and detection methods, Rept. Prog. Phys.} {\bf 63} (2000) 793.
\bibitem{Teo04}
G. Bertone, D. Hooper and J. Silk, \emph{Particle dark matter: Evidence, candidates and constraints, Phys. Rept.} {\bf 405} (2005) 279.
\bibitem{Teo01}
S. Calchi Novati, \emph{Microlensing in Galactic Halos, Il Nuovo Cimento}   {\bf B 122} (2007) 557.
\bibitem{Roc11}
R. da Rocha, A. E. Bernardini and J. M. Hoff da Silva,
\emph{Exotic dark spinor fields,
JHEP} {\bf 1104}  (2011) 110 [{\tt arXiv:1103.4759}].
\bibitem{boe1}
C. G. Bohmer,
\emph{The Einstein-Elko system -- Can dark matter drive inflation?,
Annalen Phys.} {\bf 16} (2007) 325 [{\tt {}gr-qc/0701087}].
\bibitem{boe2}C. G.  Bohmer, \emph{The Einstein-Cartan-Elko system,
Annalen Phys.} {\bf 16} (2007) 38 [{\tt {}gr-qc/0607088}].
\bibitem{boe3}C. G. Bohmer, \emph{Dark spinor inflation -- theory primer and dynamics,
Phys. Rev.} {\bf D 77} (2008) 123535 [{\tt arXiv:0804.0616}].
\bibitem{boe4} C. G. Bohmer and J. Burnett,
\emph{Dark spinors with torsion in cosmology,
Phys. Rev.} {\bf D 78} (2008) 104001 [{\tt arXiv:0809.0469}]. 
\bibitem{fabbri1}  L.~Fabbri,
  \emph{The most general cosmological dynamics for ELKO Matter Fields,
  Phys.\ Lett.} {\bf B 704} (2011) 255
  [{\tt arXiv:1011.1637}]. \bibitem{fabbri2}L.~Fabbri,
 \emph{Zero Energy of Plane-Waves for ELKOs,
  Gen.\ Rel.\ Grav.} \ {\bf 43} (2011) 1607
  [{\tt arXiv:1008.0334}]. \bibitem{fabbri3} L.~Fabbri,
  \emph{Causality for ELKOs,
  Mod.\ Phys.\ Lett.}\ {\bf A 25} (2010) 2483
  [{\tt arXiv:0911.5304}].
  \bibitem{where} R. da Rocha and W. A. Rodrigues, Jr., 
  \emph{Where are ELKO spinor fields in Lounesto spinor field classification?, 
  Mod. Phys. Lett. } {\bf A 21} (2006) 65  [{\tt {}math-ph/0506075}].
   \bibitem{where1}  R. da Rocha and J. M. Hoff da Silva,
\emph{ELKO, flagpole and flag-dipole spinor fields, and the instanton Hopf fibration,
Adv.\ Appl.\ Clifford Alg.}, {\bf 20} (2010) 847 [{\tt arXiv:0811.2717}].
 \bibitem{where2}  R. da Rocha, A. Bernardini, J. Hoff da Silva, 
\emph{ELKO Spinor fields as a tool for probing exotic topological space-time features,
 Int. J.  Mod. Phys.: Conf. Ser.} {\bf 3} (2011) 133.
  \bibitem{where3}  R. da Rocha and J. M. Hoff da Silva, 
  \emph{ELKO Spinor Fields: Lagrangians for Gravity derived from Supergravity,
 Int. J. Geom. Meth. Mod. Phys.} {\bf 6} (2009) 461 [{\tt 	arXiv:0901.0883}].
\bibitem {allu}
D. V. Ahluwalia-Khalilova and D. Grumiller,
\emph{Spin Half Fermions, with Mass Dimension One: Theory, Phenomenology, and Dark Matter},
\emph{JCAP} {\bf 07} (2005) 012 [{\tt {}hep-th/0412080}]. 
\bibitem {allu1} D. V. Ahluwalia-Khalilova and D. Grumiller,
\emph{Dark matter: A spin one half fermion field with mass dimension one?,
Phys. Rev. }{\bf D 72} (2005) 067701 [{\tt {}hep-th/0410192}].
\bibitem {allu2}D. V. Ahluwalia-Khalilova,
\emph{Theory of neutral particles: Mclennan-Case construct for neutrino, its generalization, and a new wave equation,
Int. J. Mod. Phys. }{\bf A 11} (1996) 1855 [{\tt {}hep-th/9409134}].\bibitem {allu3} D. V. Ahluwalia-Khalilova,
\emph{Dark matter, and its darkness,
Int. J. Mod. Phys. }{\bf D 15} (2006) 2267 [{\tt {}hep-th/0603545}].
\bibitem{alex} A. E. Bernardini and R. da Rocha,
\emph{Lorentz-violating dilatations in the momentum space and some extensions on non-linear actions of Lorentz algebra-preserving systems,
Phys. Rev.} {\bf D 75} (2007) 065014 [{\tt {}hep-th/0701094}].
\bibitem{alex1}A. E. Bernardini and R. da Rocha,
\emph{Obtaining the equation of motion for a fermionic particle in a generalized Lorentz-violating system framework,
Europhys. Lett.} {\bf 81} (2008) 40010 [{\tt {}hep-th/0701092}].
\bibitem{horv1}
D. V. Ahluwalia and S. P. Horvath,
\emph{Very special relativity as relativity of dark matter: the Elko connection,
JHEP} {\bf 1011} (2010)  078 	[{\tt arXiv:1008.0436}].
\bibitem{glashow}
A. G. Cohen and S. L. Glashow, \emph{Very Special Relativity, Phys. Rev. Lett.} {\bf 97} (2006) 021601 [{\tt {}hep-ph/0601236}]. \bibitem{glashow1} A. G. Cohen and S. L. Glashow, \emph{A Lorentz-Violating Origin of Neutrino Mass?} [{\tt {}hep-ph/0605036}].
\bibitem{osmano} R. da Rocha and J. M. Hoff da Silva,
\emph{From Dirac spinor fields to eigenspinoren des ladungskonjugationsoperators,
J. Math. Phys. }{\bf 48} (2007) 123517  [{\tt arXiv:0711.1103}]. 
\bibitem{osmano1} J. M. Hoff da Silva and R. da Rocha,
\emph{From Dirac Action to ELKO Action,
Int. J. Mod. Phys.} {\bf A 24} (2009) 3227 [{\tt arXiv:0903.2815}].
\bibitem{Asselmeyer:1995jp}
G. Hess, {Exotic Majorana spinors in (3+1)-dimensional space-times}, 
\emph{J. Math. Phys.} {\bf 35} (1994) 4848. 
\bibitem{isham1}
S. J. Avis and C. J. Isham,
\emph{Lorentz gauge invariant vacuum functionals for quantized spinor fields in non-simply connected space-times,
Nucl. Phys.}  {\bf B 156} (1979) 441.
\bibitem{isham2} C. J. Isham,
\emph{Twisted quantum fields in a curved space-time,
Proc. R. Soc. London, Ser.} {\bf A 362} (1978) 383.
\bibitem{isham3}C. J. Isham,
 \emph{Spinor fields in four-dimensional space-time,
Proc. R. Soc. London, Ser.} {\bf A 364} (1978) 591.\bibitem{isham4}
H. R. Petry,
\emph{Exotic spinors in superconductivity,
J. Math. Phys.} {\bf 20} (1979) 231.
\bibitem{isham5}
A. Chockalingham and C. J. Isham, \emph{Twisted supermultiplets,
J. Phys.} {\bf  A 13} (1980) 2723.
\bibitem{Ber08A}
A. E.  Bernardini and O. Bertolami, \emph{Stationary condition in a perturbative approach for mass varying neutrinos, Phys. Lett.} {\bf B 662} (2008) 97 [{\tt 	arXiv:0802.4449}]. \bibitem{Ber08B}
A. E.  Bernardini and O. Bertolami, \emph{Lorentz violating extension of the Standard Model and the $\beta$-decay end-point, Phys. Rev.} {\bf D 77} (2008) 083506 [{\tt 	arXiv:0802.2199}].
\bibitem{Ber08C}A. E.  Bernardini and O. Bertolami, \emph{Stability of mass varying particle lumps, Phys. Rev.} {\bf D 80} (2009) 123011  [{\tt 	arXiv:0909.1541}]. \bibitem{Ber08D}A. E.  Bernardini and O. Bertolami, \emph{Coupling active and sterile neutrinos in the \emph{cosmon} plus \emph{seesaw} framework, Phys. Lett.} {\bf B 684} (2010) 96 [{\tt 	arXiv:0911.0446}].
\bibitem{marcao}
M. Dias, F. de Campos and J. M. Hoff da Silva,
\emph{Searching for $\mu^{+}\mu^{-}+$ two light Elkos signal at the LHC, Phys. Lett.} {\bf B 706} (2012) 352
[{\tt arXiv:1012.4642}].
\bibitem{local} 
  D.~V.~Ahluwalia, C.~-Y.~Lee and D.~Schritt, \emph{Self-interacting Elko dark matter with an axis of locality,
  Phys.\ Rev.}\ {\bf D 83} (2011) 065017
  [{\tt arXiv:0911.2947}]. \bibitem{local1}   D.~V.~Ahluwalia, C.~-Y.~Lee, D.~Schritt and T.~F.~Watson,
\emph{Elko as self-interacting fermionic dark matter with axis of locality,
  Phys.\ Lett.\ } {\bf B 687} (2010) 248 
  [{\tt arXiv:0804.1854}].
  \bibitem{ture} M. M. Sheikh-Jabbari and A. Tureanu, \emph{Realization of Cohen-Glashow Very Special Relativity on Noncommutative Space-Time, 
  Phys. Rev. Lett.} {\bf 101} (2008) 261601  [{\tt 	arXiv:0806.3699}].   \bibitem{ture1} R. da Rocha, A. E. Bernardini, and J. Vaz Jr, 
  \emph{$\kappa$-Deformed Poincar\'e Algebras and Quantum Clifford-Hopf Algebras, 
  Int. J. Geom. Meth. Mod. Phys.} {\bf 7} (2010) 821 	[{\tt arXiv:0801.4647}].
\bibitem{clo}  D. V. Ahluwalia, \emph{Towards a relativity of dark-matter rods and clocks, 
Int.\ J.\ Mod.\ Phys.\ } {\bf D 18} (2009) 2311
 [{\tt arXiv:0904.0066}].



\end{thebibliography}
\end{document}